\documentclass[twocolumn,superscriptaddress,prl,
aps,psfig]{revtex4} 

\newcommand{\Litwo}{Li$_2$CuO$_2$}
\newcommand{\Catwo}{Ca$_2$Y$_2$Cu$_5$O$_{10}$}
\newcommand{\Livcu}{LiVCuO$_{4}$}
\newcommand{\Licu}{LiCu$_2$O$_2$}
\newcommand{\Nacu}{NaCu$_2$O$_2$}
\newcommand{\Lizru}{Li$_2$ZrCuO$_{4}$}

\newcommand{\mb}{$\mu_{\mbox{\tiny B}}$}
\usepackage{graphicx,color,epsfig,rotate}

\begin{document}

\title{Quasi-1d quantum helimagnets: The fate of
multipolar phases}

\author{S.~Nishimoto}

\author{S.-L.~Drechsler$^{*}$}

\author{R.\ Kuzian}

\affiliation{IFW Dresden, P.O.~Box 270116, D-01171 Dresden,
Germany}
\author{J.\ Richter}
\affiliation{Universit\"at Magdeburg, Institut f\"ur Theoretische Physik, Germany}
\author{J.\ van den Brink }
\affiliation{IFW Dresden, P.O.~Box 270116, D-01171 Dresden,
Germany}

\date{\today}
\begin{abstract}
Coupled frustrated spin-1/2 chains
in high magnetic fields described within the  ferro- antiferromagnetic $J_1$-$J_2$ Heisenberg model
are studied by DMRG, hard core boson, and spin wave theory approaches.
Multipolar phases related to magnon bound
states  are destroyed (supported) by  weak antiferromagnetic (ferromagnetic) interchain couplings $J_{\mbox{\tiny ic}}$. 
We show that
quantum spin nematics might be found for LiVCuO$_4$ whereas for Li(Na)Cu$_2$O$_2$ 
it is prevented by a sizeable antiferromagnetic $J_{\mbox{\tiny ic}}$. Also for Li$_2$ZrCuO$_4$ 
with a small antiferromagnetic $J_{\mbox{\tiny ic}}$ expected  triatic or quartic phases  are
unlikely, too.
The saturation field is found to be 
strongly affected even by  a relatively small $J_{\mbox{\tiny ic}}$.

\end{abstract}


\maketitle

The frustration of magnetic interactions allows for 
new phenomena to emerge.
Some spin-chain compounds  are frustrated
 magnets in, for instance, the case of ferromagnetic (FM) nearest neighbor (NN) exchange interaction
 ($J_1$) 
competing with an antiferromagnetic (AFM) 
next-neighbor (NNN) one ($J_2$)~\cite{Drechsler05}. Recent theoretical studies indicate 
that in a high magnetic field, such a $J_1$-$J_2$ chain can display e.g.\ nematic quasi 
long-range order for certain $\alpha $=$-J_2/J_1$~\cite{Kecke07,Hikihara08,Sudan09}. This nematic state might be 
thought of as a condensate of two-magnon bound states \cite{Zhitomirsky10}. Depending on the value of
 $\alpha$  also three- or four- magnon boundstates might condense, 
resulting in a very rich phase diagram with exotic magnetic multipolar (MMP) 
phases~. However, to establish whether these MMP phases can be 
realized in a real material it is essential to determine their 
robustness, i.e.\ the very existence of multimagnon bound states (MBS) with respect to 
various 
interchain (IC) couplings $J_{\mbox{\tiny ic}}$, which can be 
(very) small for certain spin-chain compounds, but that is always 
present, and, as we will show here can be detrimental.
Such a transition from 1D to 2D or 3D can be non-trivial. 
In particular, from quantum mechanics it is well-known that bound states are 
strongly pronounced in 1D and more rare in 2D or 3D.
To understand {\it quantitatively} 
the role of IC
the question
 arises in which cases 
 even a relatively weak IC 
is still important or even 
crucial? 
Here we address such a problem, 
namely, the fate of MMP states,
such as the nematic spin state 
related to MBS
at high magnetic 
fields and 
consider the 
spin-1/2 frustrated 
isotropic $J_1$-$J_2$-Heisenberg 
model mentioned above supplemented by various IC \cite{Hamer09}.
Recently, 
the 1D-model and related compounds revealed considerable interest  
\cite{Bursill95,Vekua07,Heidrich-Meisner08,Drechsler05,
Dmitriev09,
Drechsler07,Schmitt09,Sato09,Lorenz09,Zinke09,Nishimoto2010,Buettgen10,Svistov10,
Kuzian08,haertel08,
Sirker10}.
 Nowadays it is 
 the most popular
model 
for edge-shared chain cuprates (see e.g.\ \cite{Drechsler05}).
Additional AFM couplings
typically
provided by the IC
enhance the kinetic energy of magnons. 
Hence, 
AFM(FM) IC  may hinder(favor)
the formation of
low-lying MBS.
A FM IC may 
create
even new  MMP states.
An
examination of 
real
Q-1D systems to predict
the changes due to finite $J_{\mbox{\tiny ic}}$ and
to evaluate  
the chances
to detect MMP states
in certain compounds 
is of broad interest 
\cite{Katsura08,Lorenz09,Zinke09,Ueda09,Zhitomirsky10,Nishimoto2010,
Buettgen10,
Svistov10,Enderle05,Enderle10}.
\begin{figure}[b!]
\includegraphics[width=7.6cm]{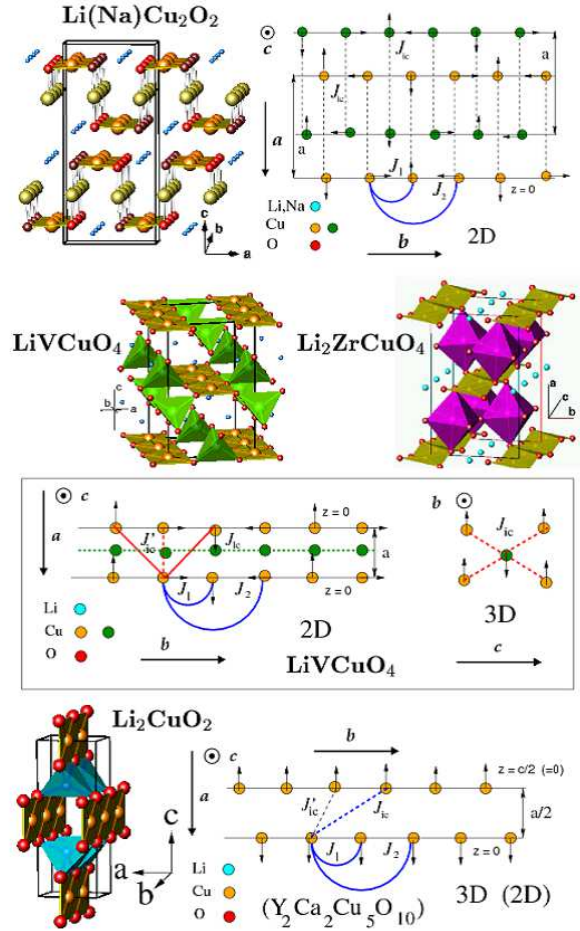}
\caption{(Color) 
Crystal structure and exchange patterns of chain cuprates. 
The main
 in- and interchain exchange 
paths
are marked by
arcs and lines, respectively. 
Upper: 1/2-unit cell projection 
(right). Planes with green and yellow Cu sites are treated as uncoupled.
We take into account
only perpendicular inplane $J_{\mbox{\tiny ic}}$.
Middle: The main 2D IC in the basal plane of unshifted chains (left) and with 3D IC (right).
Lower: Two NN shifted chains in different ($ab$)-planes. 
We ignore the weak $J'_{\mbox{\tiny ic}}$ and keep
only the second diagonal $J_{\mbox{\tiny ic}}$ \cite{Lorenz09}.
} 
\label{fig::Struct}
\end{figure}

Among edge-shared cuprates
 \Licu , the isomorphic \Nacu , and 
\Livcu $\equiv$LiCuVO$_4$ \cite{remchem} (see Fig.~\ref{fig::Struct}; for comparison the 
reference systems \Litwo \ and \Catwo \ with strong IC are shown, too) 
have been proposed to be  candidates
for quantum-spin nematics, i.e.\
quadrupolar  phases derived from 2-MBS.
Systems 
like \Lizru \ \cite{Drechsler07,Sirker10} 
located 
closer to the FM-spiral
critical point might be candidates 
for the triatic or quartic MMP. 
Besides the strength also 
the influence of various IC topologies resulting from different
arrangements of individual
chains is of interest. Various types of IC are shown in 
Fig.~\ref{fig::Struct}. The simplest case is given by unshifted neighboring
chains and a predominant perpendicular $J^{\tiny \perp}_{\mbox{\tiny ic}}$. 
Here
spirals on NN chains are only weakly affected by an AFM IC
\cite{Zinke09} (classically their pitch angle, i.e.\ the incommensurate (INC) inchain
magnetic structure 
in the dipolar phase at ambient external fields 
remains even unaffected).
Hereafter, we call this 
IC the 'unfrustrated IC'. 
An effective 2D arrangement of the magnetically active Cu$^{2+}$-sites is realized approximately for
Li(Na)Cu$_2$O$_2$,
where
$J_{ic}\sim$(0.5 to 1)$J_2 \sim$ 40 to 100~K
\cite{Gippius04,Masuda05, Drechsler06}.
A square lattice of unshifted chains, i.e.\ the
3D case considered in
Ref.\ \onlinecite{Ueda09} might be realized
	in LiVCuO$_4$, if the IC denoted as $J_3$ in Ref.\ \onlinecite{Enderle05}
	is dominant as compared with the IC in the ($ab$)-plane (see middle of Fig.\ 1).
Another 3D case but with shifted adjacent chains 
is realized for \Litwo \ (see Fig.\ 1) for which 
$H_{\mbox{\tiny s}}$ 
has been found recently \cite{Nishimoto2010}.
$J_{\mbox{\tiny ic}}\approx $10~K 
can be extracted from
$H_{\mbox{\tiny s}}(0)$,
from inelastic neutron scattering  studies,  
and from 
bandstructure results
\cite{Lorenz09}.
The same order
holds also for Li$_2$ZrCuO$_4$ 
where buckling of the CuO$_2$ chains  reduces  $J_{\mbox{\tiny ic}}$ \cite{Schmitt09}.
\begin{figure}[t]
\includegraphics[width=8.3cm]{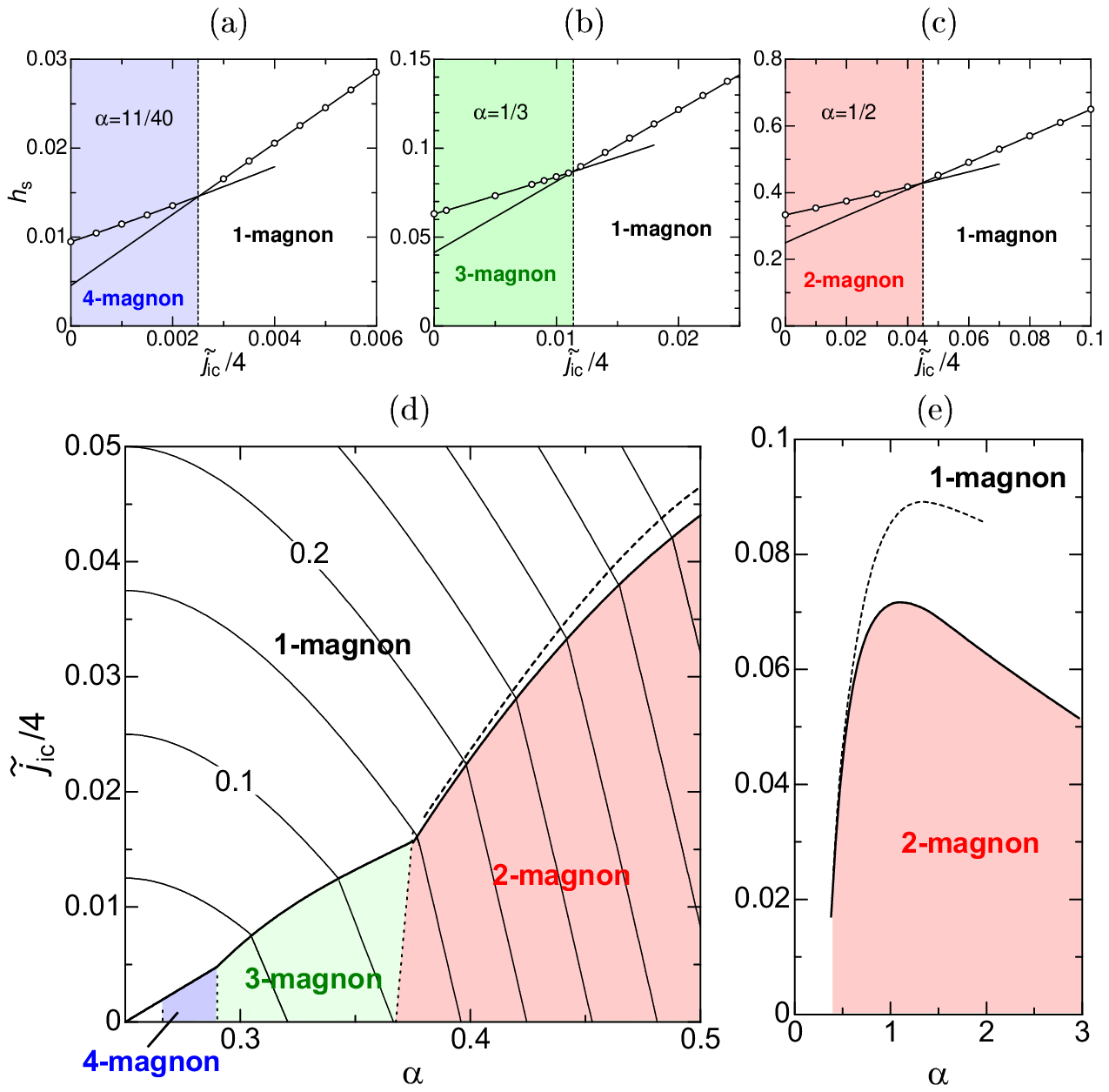}
\caption{(Color) Upper (a-c): Saturation field vs IC; lower: critical IC
for the 3D case with perpendicular IC
$\tilde{j}_{\mbox{\tiny ic}}= N_{\mbox{\tiny ic}}J^{\tiny \perp}_{\mbox{\tiny ic }}/| J_1| $ 
vs.\ inchain frustration $\alpha$ (d,e), where $N_{\mbox{\tiny ic}}=4$ is the number 
of nearest interchain neighbors. The pase boundaries are constructed from the 'kinks'
in $h_{\mbox{\tiny s}}$-plots
like in (a-c).
The contour lines in (d) show the values of
$h_{\mbox{\tiny s}}$. 
The 2D case with $N_{\mbox{\tiny ic}}$=2 is shown also in (d,e) (dotted lines).
} 
\label{f2}
\end{figure}

We used the density-matrix renormalization group (DMRG) method~\cite{White92} 
with 
imposing
periodic boundary conditions (PBC) for all directions. 
In general, it is known that this method is much less appropriate for $D>1$. 
However, spin systems with up to about 
$\sqrt{n} \times \sqrt{n} \times L = \sqrt{10} \times \sqrt{10} \times 50$ 
sites can be studied by taking a proper construction of the lattice block 
(see Ref.~\onlinecite{Nishimoto2010}). We kept $m \approx 800-4000$ density-matrix 
eigenstates in the renormalization procedure. In fact, about $100-300$ sweeps 
are necessary to obtain the GS energy within a convergence of 
$10^{-7}J_1$ for each $m$ value. All calculated quantities were extrapolated 
to $m \to \infty$ and the maximum error in the GS energy is estimated 
as $\Delta E/J_1 \sim 10^{-4}$, while the discarded weight is less than 
$1 \times 10^{-6}$. Under the PBC, a uniform distribution of 
$\left\langle S^z_i \right\rangle$ may give an indication to examine 
the accuracy of DMRG calculations for spin systems. Typically, 
$\left\langle S^z \right\rangle-S_{\rm tot}^z/(nL)$ is less than $1 \times 10^{-3}$ 
in our calculations. Note that for high-spin states 
[$S_{\rm tot}^z \gtrsim (nL-10)/2)$] the GS energy can be obtained with 
an accuracy of $\Delta E/J_1 < 10^{-12}$ by carrying out several thousands 
sweeps even with $m \approx 100-800$. 
Furthermore, we studied the systems with several lengths: 
$L=16-64$ ($24-96$) for 3D (2D) and then adopting power laws we
performed a 
finite-size-scaling analysis to obtain the saturation field 
$h_{\mbox{\tiny s}}\equiv g\mbox{\mb }H_{\mbox{\tiny s}}/| J_1 |$
in the thermodynamic limit $L \to \infty$.
As a result, we obtain  $h_{\mbox{\tiny s}} $ with high accuracy.
In addition to DMRG we have also applied the linear spin wave theory and the hard core
boson approach \cite{Kuzian08}. The latter provides exact results for the 
nematic phase.
Some results for the reported magnetization curves have been additionally
checked applying the exact diagonalization. 

 We start with the simplest IC case
 of parallel chains within a plane  (2D)
 or within a
  square-lattice arrangement of chains 
  (3D)
  and 
 a single perpendicular IC $J^{\tiny \perp }_{\mbox{\tiny ic}}$=$j^{\tiny \perp}_{\mbox{\tiny ic}}|J_1|$
 (see Figs.\ 1,2).
 A typical magnetization curve 
 for 
 the nematic
 phase is shown in Fig.\ 3. The height of the magnetization steps $\Delta S^z= 2$, only,
 is its signature whereas in the 1-magnon phase $\Delta S^z =1$ holds.  
\begin{figure}[b]
\includegraphics[width=5.7cm]{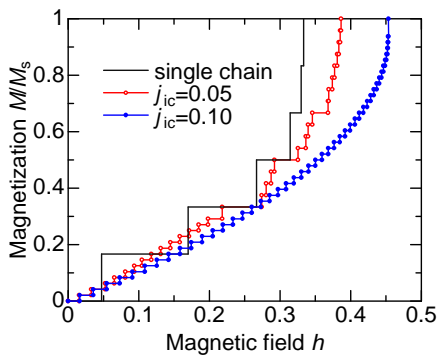}
\caption{(Color) Magnetization vs.\ 
 applied external field
for a 2D arrangement of 
chains with a direct AFM IC as modelled
by four chains with $N=24$ sites in each
chain for different 
IC 
$j_{\mbox{\tiny ic}}\equiv J_{\mbox{\tiny ic}}/ |J_1 | $ and an
 inchain frustration $\alpha=1/2$. 
 Note that each point shown represents a step.
 The critical value of $\tilde{j}_{\mbox{\tiny ic}}/2$ 
amounts 
 0.088 for $D$=3 and 0.094 for $D$=2.
}
\label{magnetization}
\end{figure}
 Note that a rather weak critical
 IC of a few percent
 will remove
 the nematic phase in favor of the usual field induced 
 "conic" phase. The 
 saturation field $h_s$ of the INC
 phase
 on the 1-magnon side is described exactly already
 within spin wave theory: 
\begin{equation}
h_s\equiv g\mu_{\mbox{\tiny B}}H_{\mbox{\tiny s}}/| J_1| =
2\alpha -1+0.125/\alpha  +\tilde{j}_{\mbox{\tiny ic}}  ,
\end{equation}
where $\tilde{j}_{\mbox{\tiny ic}}$=$N_{\mbox{\tiny ic}}j_{\mbox{\tiny ic}}$=$2(D-1)j_{\mbox{\tiny ic}}$ 
for $D$=2,3.
Using the Curie-Weiss temperature $\Theta_{\mbox{\tiny CW}}= 
 -(J_1+J_2+0.5N_{\mbox{\tiny ic}}J_{\mbox{\tiny ic}} )/2$
which determines the high-temperature
spin susceptibility $\chi(T)\sim 1/(T-\Theta_{\mbox{\tiny CW}})$, 
the IC in Eq.\ (1)
can be eliminated. 
Then in this
1-magnon phase one has a simple expression 
 to
extract  $J_1$ or $\alpha$ from $H_s$ and $\Theta_{\mbox{\tiny CW}}$:
\begin{equation}
g\mbox{\mb}H_{\mbox{\tiny s}}(J_{\mbox{\tiny ic}})+4\Theta_{\mbox{\tiny CW}}(J_{\mbox{\tiny ic}})\equiv G =
|J_1| \left[1+1/(8\alpha )\right].
\label{empiric}
\end{equation}
A relation $G= |J_1| f(\alpha )$ is valid also in other regimes, where $f$ is
a simple function affected by the type of the IC and the region of the MMP phase diagram.
In the 1-magnon regimes  valid for \Litwo \ and Ca$_2$Y$_2$Cu$_5$O$_{10}$ (see below) 
we have  $f=2(1-\alpha )$ \cite{Nishimoto2010}.
For the nematic phase in full
accord with the DMRG
we obtained numerically
{\it exact} values of $h_{\mbox{\tiny s}}$ using   
the hard-core boson approach \cite{Kuzian08}.
 Expanding the 2-particle Green's function in powers of $j_{\mbox{\tiny ic}} \ll 1$ 
we arrive at analytical expressions:
\begin{equation}
h_{\mbox{\tiny s}}(\alpha )\approx h^{\mbox{\tiny 1D}}_{\mbox{\tiny s}}(\alpha )+
\tilde{j}_{\mbox{\tiny ic}}/2 +N_{\mbox{\tiny ic}}\eta ( \alpha ) j^2_{\mbox{\tiny ic}} 
+O\left( j^4_{\mbox{\tiny ic}} \right),  
\label{hs2mag}
\end{equation}
\vspace{-0.6cm}
\begin{equation}
h^{\mbox{\tiny 1D}}_{\mbox{\tiny s}}(\alpha )= 2\alpha -1 +0.5/(1+\alpha ),
\end{equation}
where 
$\eta (\alpha)
=(1+\alpha)(3\alpha^2+3\alpha+1)/[2(1+2\alpha)^2]\approx 5/3+3\alpha /2 $ . 
The expansion coefficients for higher order terms 
differ for  $D$=2 and 3 reflecting subtle natural differences 
in the quantum fluctuations. A detailed discussion will be given elsewhere.
Comparing the Eqs.\ (1) and (3) we stress the presence of nonlinear IC terms  and two times 
smaller linear term in the nematic phase as compared with the usual one-magnon phase.
The inspection of Figs.\ 2(d,e) shows that the critical IC of the nematic 
 region
 reaches a maximum
 as a function of 
 $\alpha$.
   Its position  $\alpha \approx 1.103$ and height $\tilde j_{ic}/4 \approx 0.071688$
 are close to the
 feature shown in Fig.\ 6 of Ref.\ \onlinecite{Ueda09}.

Keeping the linear IC term in Eq.\ (\ref{hs2mag}) we arrive at 
\begin{equation}
g\mbox{\mb}H_{\mbox{\tiny s}}+2\Theta_{\mbox{\tiny CW}}\equiv K=| J_1| \left[\alpha +0.5/(1+\alpha )\right].
\end{equation}
Thus, $G$ and $K$ composed from IC affected quantities depends itself on single chain properties, only.
In most cases it is
easier to determine the latter theoretically whereas $G$ or $K$ can be found from experiment. 

Applying Eq.\ (\ref{empiric}) to Li$_2$ZrCuO$_4$ we predict $\Theta_{\mbox{\tiny CW}}=93.3$~K
for a preliminary experimental value of $H_s$
of about 13~T \cite{Tristan}
using $J_1=273$~K,
$\alpha=0.29$ to 0.3 
 \cite{Drechsler07}. The small 1D-value of  
 $H_{\mbox{\tiny s}} \approx 4.2$ to 5.9~T, only,
 clearly shows the importance of $J_{\mbox{\tiny ic}}$ in the vicinity of the 
 critical point where $H_s \rightarrow 0$. The validity of Eq.\ (\ref{empiric})
 is guaranteed
 by $j_{\mbox{\tiny ic}}=$ 0.0264756 (i.e. 7.2~K) to 0.023659  (i.e.\ 6.5~K) exceeding well
 $j_{\mbox{\tiny ic,cr}}$=0.009539 
 to 0.013458 deduced from the
  2D-phase diagram.
 The $j_{\mbox{\tiny ic,cr}}$
 would allow $H_{\mbox{\tiny s}}=6.2$~T to 8.9~T 
 in the triatic phase
 at most. The estimated weak IC given above is close to   L(S)DA+$U$ results \cite{Schmitt09}.
 Thus, we are clearly outside the triatic or quartic phases  and deep enough in 
 the usual
 dipolar phase. The 1-magnon 
 picture holds also for Li(Na)Cu$_2$O$_2$ where $j_{\mbox{\tiny ic,cr}}$
 for the
 nematic phase
 is exceeded
 by a factor of four. In fact, from the LDA derived $J_{\mbox{\tiny ic}}$ and $J_1$ we estimate 
 $j_{\mbox{\tiny ic}} \approx 0.7$ well above 0.170 
 $j_{\mbox{\tiny ic,cr}}=0.1704$
 taken 
 from Fig.\ 2 for $\alpha
 \approx 1$ for \Licu \cite{Gippius04}.
 Similarly, using the empirical values for \Nacu : 
 $\Theta_{\mbox{\tiny CW}}=-41$~K taken from the $1/\chi(T)$ data at 350 $\leq T \leq$
  400~K, 
 $\alpha=1.9$, $J_1\approx-48~$K, and $g_a=2.06$,
 we predict  $H_s \sim$ 155~T. From Eq.\ (1) we estimate 
 $j_{\mbox{\tiny ic}} \sim $ 0.73  well above
 $j_{\mbox{\tiny ic,cr}}\approx $ 0.172 (see
 Fig.\ 2) and again close to LDA-results \cite{Drechsler06}.

Now we turn to the case of 
diagonal IC as in \Litwo \ (see Fig.\ 1).
\begin{figure}[b]
\includegraphics[width=5.8cm]{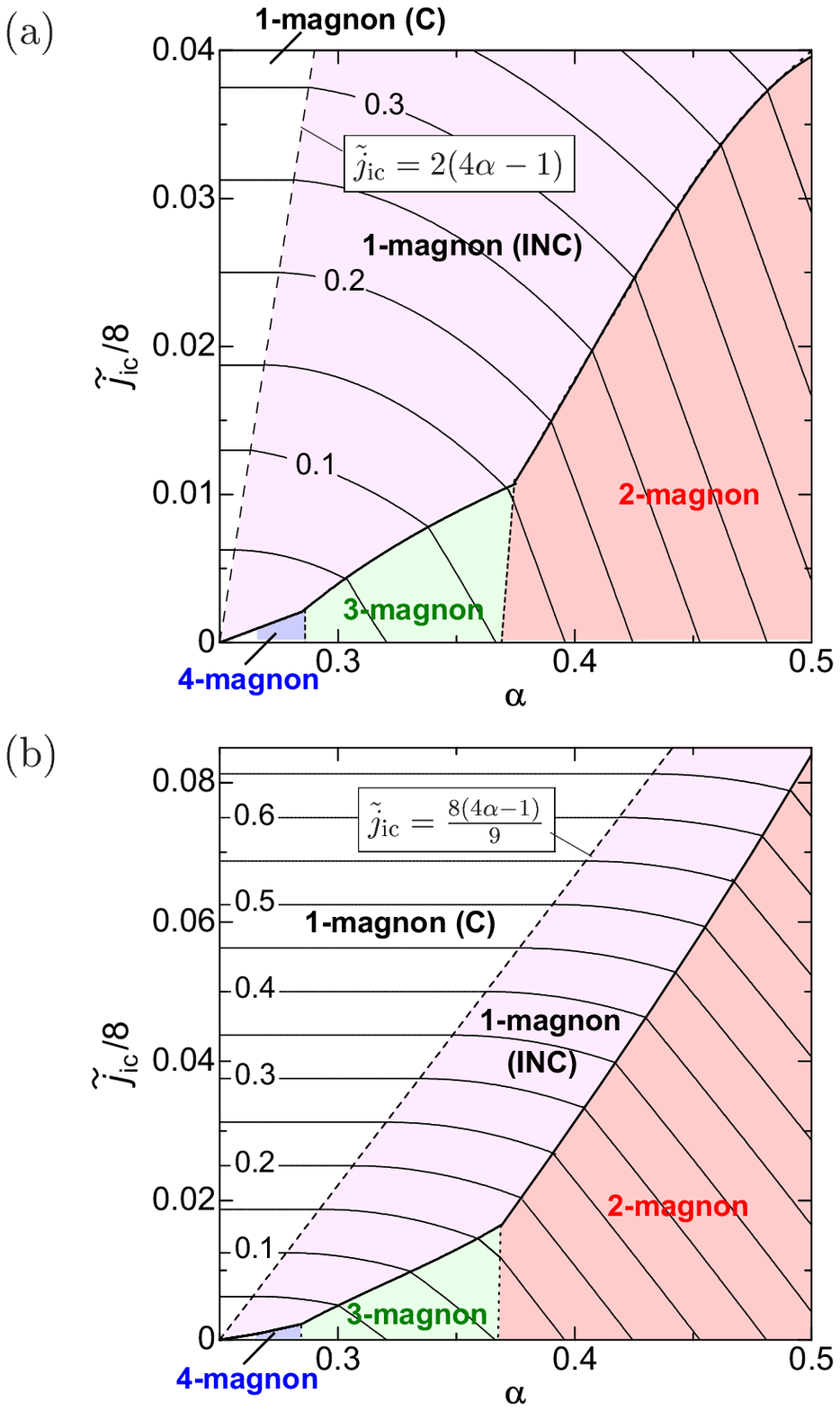}
\caption{(Color)
3D phase diagrams for diagonal IC.
$h_s$ (given by contour lines)
as a function of the normalized
IC $\tilde{j}_{\mbox{\tiny ic}}$ and the inchain frustration $\alpha$.
Lower(Upper): (un)shifted chains with $\pm 3b/2$ ($\pm b$) IC, where $b$ is the inchain Cu-Cu distance.
} 
\label{f5}
\end{figure}
For a strong enough IC the
MBS are 
supressed and only 1-magnon excitations survive
at high fields (see Fig.\ 4(b)).
Then $h_s$ reads
\begin{equation}
h_{\mbox{\tiny s}}=g\mu_{\mbox{\tiny B}} H_{\mbox{\tiny s}}/|J_1| =  
\tilde{j}_{\mbox{\tiny ic}}+\tilde{j}'_{\mbox{\tiny ic}}
\ \mbox{if} \ j_{\mbox{\tiny ic}} \geq j^{\mbox{\tiny cr}}_{\mbox{\tiny ic,1}},\ 
\tilde{j}_{\mbox{\tiny ic}}=N_{\mbox{\tiny ic}}j_{\mbox{\tiny ic}}
\label{hsatli}
\end{equation}
where   $N_{\mbox{\tiny ic}}= 8$.
Eq.\ (\ref{hsatli}) is valid for 
$j_{ic}(\alpha) > (4\alpha -1)/9\equiv j^{\mbox{\tiny cr}}_{\mbox{\tiny ic,1}}$
for $j'_{\mbox{\tiny ic}}=0$.
We obtained from INS
data \cite{Lorenz09}:
$J_{ic}\approx 9 \mbox{K} > J^{\mbox{\tiny cr}}_{\mbox{\tiny ic,1}}(0.332)=-0.0364J_1 \approx $8.2~K.
Then $H_{\mbox{\tiny s}}$
 depends {\it solely} on the IC which
can be directly read off from 
$H_{\mbox{\tiny s}}(T=0)$ \cite{Nishimoto2010}.
 For weaker IC $h_s$ depends also somewhat on
$\alpha$. 
Our 
results 
also suggest that in this intermediate 
INC-phase
above a
second critical IC
$j^{\mbox{\tiny cr}}_{\mbox{\tiny ic,2}}\approx 0.0109$ only specific INC
1-magnon low-energy excitations  exist, see Fig.\ 4.
Below $j^{\mbox{\tiny cr}}_{\mbox{\tiny ic,2}}$ 3-MBS are recovered 
as low-energy excitations
and only
this narrow
region is 1D-like.
The commensurate (C) phase behaves like an ordinary 3D antiferromagnet despite
its seemingly
quasi-1D nature.
 For unshifted NN chains  (see Fig.\ 1),
  the C-phase is missing. There is 
only one critical IC  separating INC 1-magnon excitations 
from 3-MBS.

Generally, 
$\theta =-2\Theta_{\mbox{\tiny CW}}/J_1 $  provides a useful
constraint 
\begin{equation}
1=\alpha + \beta  +\theta +(D-1)[j_{\mbox{\tiny ic,$\perp$}}+2\sum_fj_{\mbox{\tiny ic,f }}] ,
\label{cw}
\end{equation}
for the exchange integrals,
where in addition to $\alpha$ further inchain and IC quantities 
$\beta=-J_3/J_1$, $j_{\mbox{\tiny ic,$\perp$}}=-J_{\mbox{\tiny
ic,$\perp$}}/J_1$, and 
$j_{\mbox{\tiny ic,f}}=-J_{\mbox{\tiny
ic,f}}/J_1$
for various diagonal couplings have been introduced. 
From Eq.\ (\ref{cw}) a stringent constraint for $\alpha$ for LiVCuO$_4$ follows. From 1/$\chi(T)$ at
500~K $\leq T \leq $  650~K a small {\it FM} value $\Theta_{\mbox{\tiny CW}} =7.4$~K 
can be fitted from the data  
shown in 
Refs.\ \onlinecite{Enderle05,Drechsler10}. 
The FM ordering of spirals observed by neutron diffraction can maintained both by
a FM inplane and a perpendicular AFM 3D IC (see Fig.\ 1)
Then we may assume for the sake of simplicity that both IC  almost cancel in Eq.\ (\ref{cw}) and
ignore the  small terms $\beta$, and $j_{\mbox{\tiny ic}}$ in zeroth approximation.
The weakness of the total IC is
 suggested also by the weak magnetic moment $m=0.3 \mu_{\mbox {\tiny B}}$ at $T$= 0 \cite{Gibson04}
 pointing to
 strong quantum fluctuations.   
A first estimate \cite{Drechsler10} of the dispersion of the
INS peaks   as well as of the INS intensity above 10~meV 
yields a reasonable
description 
for $\alpha $=0.8 and $J_1$=$-$73.2~K and $J_3$=$-$0.07~K as suggested by    
a mapping of a five-band Hubbard model (exact diagonalizations) 
for a Cu$_6$O$_{12}$ cluster onto a  six-site 
$J_1$-$J_2$-$J_3$-Heisenberg ring.
Hence, we 
conclude, that for $\alpha \approx 0.85 \pm 0.1$ and $J_1  \approx -80$~K nematics could be observed in
\Livcu . 
Using Eq.\ (\ref{hs2mag}) we see that this 
is also in reasonable
accord with the experimental $H_s$-data \cite{Enderle05,Svistov10} since our 1D set  
gives for the above
mentioned rough estimate about 48~T for $g=2$ and H  directed along the hard axes and about 42~T for the
weak axis with $g=2.3$.
Note that our $J$'s   differ strongly from those given inappropriately
in Refs.\ {\onlinecite{Enderle05,Enderle10}}. The latter have
been employed unfortunately in Refs.\ \onlinecite{Zhitomirsky10,Svistov10}
as 'empirical' input parameters. 
 More detailed 
studies including explicitly 
the IC are necessary to refine all $J$'s. 
In particular, the quantum effect of $J_{\mbox{\tiny ic}}$
on the pitch angle is of interest. A more detailed study of FM IC as well as of the $J_1$-$J_2$-$J_3$ inchain model 
will be considered elsewhere. A
 challenging point is also
to find MP phases {\it beyond} the 
2-magnon Bose condensation \cite{Zhitomirsky10} triggered by a FM IC.

To summarize, the crucial role of realistic AFM
IC in Q-1D helimagnets
has been demonstrated.
The rich and exotic physics of multipolar phases recently 
predicted for single chains is very sensitive to the strength
of the IC. 
It can be 
readily 
eliminated by a 
weak AFM IC especially for  triatic, quartic, etc.\ phases.
For most CuO$_2$ chain systems studied so far, except probably LiVCuO$_4$,
where a nematic phase or some other exotic phase might be expected,
 the AFM IC is  too strong 
to allow for
multipolar phases.
The examination of various weak FM IC as well as of anisotropy effects in real materials 
is under study. 

We thank the
DFG 
[grants 
DR269/3-1 (S-LD, SN), RI615/16-1 (JR) 
and 
the PICS program (Contr.\ CNRS No.\ 4767, NASU No.\
243) [ROK] for support. 


\begin{thebibliography}{99}

\bibitem[*]{dre}Corresponding author: s.l.drechsler@ifw-dresden.de
\bibitem{Drechsler05}S.-L.\ Drechsler {\it et al.}, J.\ Magnet. \& Magnet.\ Mat.\ {\bf 290}, 345 (2005),
--, ibid.\ {\bf 316}, 306 (2007). 
\bibitem{Kecke07}L.\ Kecke
{\it et al.} 
Phys.\ Rev.\ B 
{\bf 76}, 060407 (2007).
\bibitem{Hikihara08}T.\ Hikihara
{\it et al.}, ibid.\ 
{\bf 78}, 144404 (2008).
\bibitem{Sudan09}J.\ Sudan
{\it et al.}, ibid.\  
{\bf 80}, 140402(R) (2009).
\bibitem{Zhitomirsky10}
M.\ Zhitomirsky and 
H.\ Tsunetsugu, 
arXiv:1003.4096v1.
\bibitem{Hamer09} MBS appear at $H=$0
also in Ising-type anisotropic 2D systems, ladders 
and other  systems. See e.g.: 
C.J.\ Hamer, Phys.\ Rev.\ B {\bf 79}, 212413 (2009), S.\ Dusuel {\it et al.}, 
ibid.\ 
{\bf 81}, 064412 (2010)
and references therein. Related aspects for frustrated chain cuprates will be considered elsewhere.
\bibitem{Bursill95}R.\ Bursill
{\it et al.}, 
J.\ Phys.: Cond.\  Mat.\ 
{\bf 7}, 8605 (1995).
\bibitem{Vekua07}T. Vekua {\it et al.}, 
Phys.\ Rev.\ B {\bf 76}, 174420 (2007). 
\bibitem{Heidrich-Meisner08}F.~Heidrich-Meisner {\it et al.}, 
ibid.\ 
{\bf 74}, 020403R (2008).

\bibitem{Dmitriev09}D.~Dmitriev 
{\it et al.},  
Phys.\ Rev.\  B
{\bf 79}, 054421 (2009).
\bibitem{Drechsler07}S.-L.\ Drechsler {\it et al.}, Phys.\ Rev.\ Lett.\ {\bf 98}, 077202 
(2007).
\bibitem{Sato09}M.\ Sato {\it et al.}, Phys.\ Rev.\ B {\bf 79}, 060406(R) (2009).
\bibitem{Schmitt09}M.\ Schmitt
{\it et al.}, ibid.\ , 
{\bf 80}, 205111 (2009).
\bibitem{Kuzian08}R.~Kuzian and S.-L.\ Drechsler, 
ibid.\ 
{\bf 75}, 024401 (2007).

\bibitem {Zinke09} 
R.\ Zinke
 {\it et al.}, ibid.\, 
{\bf 79}, 094425  (2009). 
\bibitem{Lorenz09}W.E.A.\ Lorenz {\it et al.}, 
Europhys.\ Lett.\ {\bf 88}, 37002 (2009).
\bibitem{Nishimoto2010}S.\ Nishimoto {\it et al.}, 
arXiv:1004.3300 (2010).
\bibitem{Buettgen10}N.\ Buettgen {\it et al.}, Phys.\ Rev.\ B {\bf 81}, 052403 (2010).
\bibitem{Svistov10}L.E.\ Svistov {\it et al.}, arXiv:1005.5668v1
\bibitem{haertel08}M.~H\"{a}rtel 
{\it et al.},
Phys. Rev. B 
{\bf 78}, 174412 (2008).


\bibitem{Sirker10}J.\ Sirker,
ibid.\ {\bf 81}, 014419 (2010).

\bibitem{Katsura08}H.\ Katsura {\it et al.}, Phys.\ Rev.\ Lett.\ {\bf 101}, 187207 (2008).

\bibitem{Ueda09}H.T.\ Ueda  and K.\ Totsuka, Phys.\ Rev.\ B {\bf 80}, 014417 (2009).
Here $J^{\tiny \perp}_{\mbox{\tiny ic}}/J_2 $ vs.\ -1/$\alpha$ has been
plotted;
note that in this plot its maximum occurs at $-1/\alpha \approx $
 -1.67).  
\bibitem{Enderle05}M.\ Enderle
 {\it et al.}, 
 Europhys.\ Lett.\ 
 {\bf 70}, 237 (2005).
 \bibitem{Enderle10}M.\ Enderle {\it et al.}, Phys.\ Rev.\ Lett.\ {\bf 104}, 237207 (2010).
 \bibitem{remchem} We prefare the first notation to underline its cuprate character
 at variance with the standard chemical notation with increasing charge of cations.
 \bibitem{Gippius04}A.A.\ Gippius
{\it et al.}, Phys.\ Rev.\ B 
{\bf 70}, 020406(R) (2004).
\bibitem{Masuda05}T.\ Masuda {\it et al.}, 
Phys.\ Rev.\ B,
{\bf 72}, 014405 (2005).
\bibitem{Drechsler06}
S.-L.\ Drechsler
 {\it et al.}, Europhys.\ Lett.\ {\bf 73}, 83
(2006).
\bibitem{White92} S.R.~White, Phys.\ Rev.\ Lett.\ {\bf 69}, 2863 (1992).


\bibitem{Tristan}Y.\ Skourski (unpublished).
\bibitem{Drechsler10}S.-L.\ Drechsler {\it et al.}, arXiv:1006.5070. 
\bibitem{Gibson04}B.J.\ Gibson {\it et al.}, Physica B {\bf 350}, Suppl.\ e253 (2004).
Note that the ordered moment for \Livcu \ is 
 three times smaller than that for \Litwo \ or Ca$_2$Y$_2$Cu$_5$O$_{10}$. 



\end{thebibliography}
\end{document}